# Boundary Layer Turbulence Index: Progress and Recent Developments

Kenneth L. Pryor
Center for Satellite Applications and Research (NOAA/NESDIS)
Camp Springs, MD

## 1. Introduction

Turbulence is a frequently occurring phenomenon in the boundary layer and may pose a threat to low-flying aircraft and aircraft during the take-off and landing phases of flight. In addition, gusty surface winds, the manifestation of turbulence in the boundary layer, may also be a hazard to ground transportation and enhance fire danger. Therefore, a numerical model-derived boundary layer turbulence product has been developed and implemented (Pryor 2007). The boundary layer turbulence index (TIBL) product is designed to assess the potential for turbulence in the lower troposphere, generated using National Center for Environmental Prediction (NCEP) Rapid Update Cycle (RUC)-2 model data. Derivation of the index algorithm was based on the following assumptions and approximations as inferred from Stull (1988) and Garratt (1992):

1. Horizontal homogeneity
2. Subsidence is negligible
3. The coordinate system is aligned with the mean wind U
4. The boundary layer is sufficiently dry such that $T_v \approx T$ and $\theta_v \approx \theta$
5. $\partial\theta/\partial z \sim \partial T/\partial z$
6. $w'\theta' \sim \partial\theta/\partial z$, $w'T' \sim \partial T/\partial z$, $w'u' \sim \partial U/\partial z$, based on local closure where $\theta$ is potential temperature, $\theta_v$ is virtual potential temperature and $T_v$ is virtual temperature.

As outlined in Pryor (2007), the index algorithm approximates boundary layer turbulent kinetic energy (TKE) by parameterizing the two most important production terms of the TKE budget equation as outlined in Stull (1988) and Garratt (1992): vertical wind shear, responsible for mechanical production, and kinematic heat flux, responsible for buoyant production. Accordingly, the TIBL product parameterizes convective sources of turbulence as well as wind shear in the lower boundary layer and near the top of the boundary layer that may also contribute to turbulence generation. Therefore, assuming horizontal homogeneity and substituting absolute temperature T for potential temperature $\theta$, boundary layer TKE can be approximated by equation 1, based on simplification of the TKE budget equations as given in Stull (1988) and Sorbjan (1989):

$$\partial e/\partial t \sim \partial V/\partial z + \partial T/\partial z \qquad (1)$$

The terms on the r.h.s. of equation 1 are considered to be the most important contributors to boundary layer turbulence. The second term on the r.h.s. represents the temperature lapse rate between the 700 and 850-mb levels.

In addition to the forecasting of boundary layer turbulence potential, the TKE product may also have implications for assessing fire danger. The Haines Index (or Lower Atmosphere Stability Index) (Haines 1988) indicates the potential for wildfire growth by measuring stability and dryness in the boundary layer. Similar to the TKE index, the Haines Index calculation of stability incorporates the temperature difference between the 700 and 850-mb levels. In addition, surface wind speed has a direct influence on the spread and intensity of wildfires. Surface wind speed is a criterion by which the National Weather Service determines the necessity of a Red Flag Warning that indicates the presence of weather conditions of particular importance that may result in extreme burning conditions.

As noted in Pryor (2007), validation for the TKE index product was conducted for selected significant wind events. Two cases of significant wind events during November 2006 and February 2007 over the Oklahoma Panhandle region were presented. The February 2007 high wind event demonstrated the effectiveness of the TIBL product in quantifying the evolution of turbulence in an evolving morning mixed layer. Apparent in this study was a strong correlation between a TKE maximum and the magnitude of peak surface wind gusts observed by Oklahoma Mesonet stations. Based on the favorable results highlighted in the case studies in Pryor (2007), the RUC TKE product should have operational utility in assessing hazards to low-flying aircraft. This paper will outline recent developments and progress with the RUC-model TKE index product. Validation effort and recent case studies of significant wind events over southeastern Idaho using observations from the Idaho National Laboratory (INL) mesonet will be presented to demonstrate the effectiveness of the turbulence product in the short-term forecasting of gusty surface winds.

## 2. Methodology

The objective of this validation effort was to qualitatively and quantitatively assess the performance of the RUC TKE product by employing classical statistical analysis of real-time data. Accordingly, this effort entailed a study of significant wind events over the Eastern Snake River Plain (ESRP) of southeastern Idaho during the winter of 2008. Significant wind events constitute wind gusts of at least 18 m/s (35 knots) and present a hazard to low-flying aircraft, ground transportation, and enhance the threat of wildfires. TKE product image data was collected for thermodynamic environments associated with 12 significant wind events that occurred within the Idaho National Laboratory (INL) mesonet domain between December 2007 and February 2008. Clawson et al. (1989) provides a detailed climatology and physiographic description of the INL as well as a description of the associated meteorological observation network. Product images were generated in Man computer Interactive Data Access System (McIDAS) by a program that reads and processes numerical weather prediction model data, performs mathematical operations to calculate turbulence risk, and overlays risk value contours on product imagery. In addition, RUC model-derived sounding data, available via website http://rucsoundings.noaa.gov/gifs/ was collected to compare product images to coincident vertical temperature and wind profiles. It was found that a three-hour forecast product image provided an optimal characterization of the thermodynamic environment over the INL prior to and during significant wind events.

Wind gusts, as recorded by National Oceanic and Atmospheric Administration (NOAA) mesonet observation stations, were measured at a height of 15 meters (50 feet) above ground level. For each significant wind event, product images were compared to RUC sounding profiles and radar profiles of wind and temperature as observed by NOAA GRID3 mesonet station. Archived NOAA mesonet observations are available via the NOAA INL Weather Center website (http://niwc.noaa.inel.gov).

At this early stage in the algorithm assessment process, it is important to consider covariance between the variables of interest: turbulence risk (expressed as a TKE index value) and surface wind gust speed. A very effective means to assess the quantitative functional relationship between the TKE index algorithm output and wind gust strength at the surface is to calculate correlation between these variables. Thus, correlation between TKE index values and observed surface wind gust velocities for the selected events were computed to assess the significance of these functional relationships. Statistical significance testing was conducted, in the manner described in Pryor and Ellrod (2004), to determine the confidence level of correlations between observed wind gust magnitude and turbulence risk values. Hence, the confidence level is intended to quantify the robustness of the correlation between turbulence risk values and wind gust magnitude.

## 3. Case Studies

### 3.1 Case 1: 19 December 2007

During the afternoon of 19 December 2007, high winds were observed by mesonet stations in the INL complex, resulting from a large pressure gradient between a strong anticyclone over the central Rocky Mountains and a strong cyclone over southwestern Canada. The afternoon (2200 UTC) RUC three-hour forecast boundary layer turbulence product indicated a maximum or “ridge” in index values over INL as well as a general increase in TKE between 2100 and 2200 UTC. Strong vertical wind shear in the cloud layer above the boundary layer resulted in the generation of turbulent eddy circulations, reflected as elevated RUC TKE index values. Strong wind gusts in excess of 40 knots were observed by INL mesonet stations near 2200 UTC. Elevated TKE values, emulated in radar sounding and RUC model profiles over the INL, indicated favorability for strong surface wind gusts due to large eddy circulation and resulting downward horizontal momentum transport.

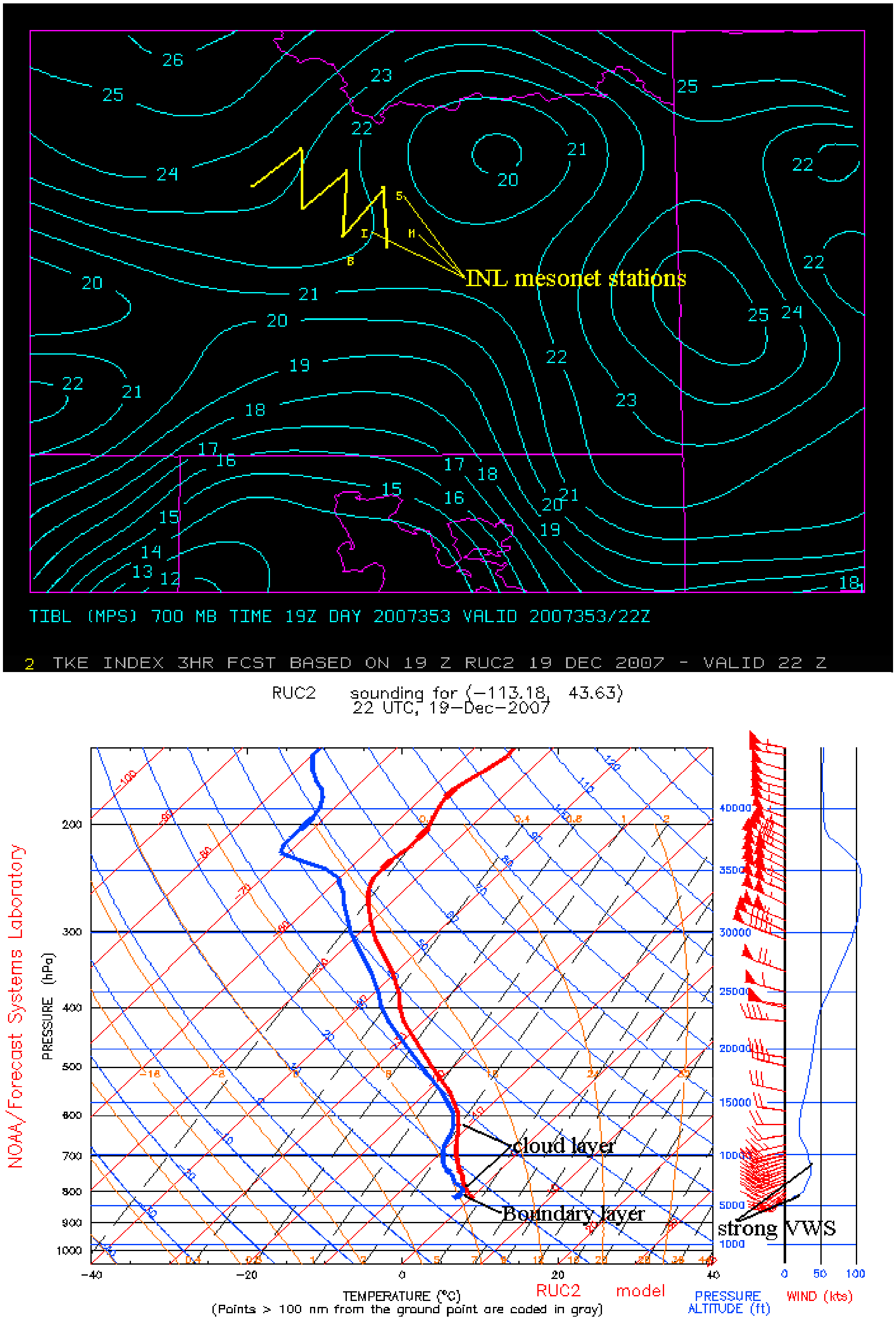

Figure 1. Turbulent Kinetic Energy (TKE) Index product image (top) and corresponding sounding profile over INL (bottom) based on the Rapid Update Cycle (RUC) model valid at 2200 UTC 19 December 2007.

The TKE image in Figure 1 displays a maximum in index values (yellow “zigzag” line) over INL. Surface observations at Critical Infrastructure Complex mesonet station (near location marked “I” in TKE image) indicated an increase in wind gust magnitude between 2140 and 2210 UTC (not shown), with a peak wind gust of 42 knots recorded at 2210 UTC. The sounding displays in Figure 1 strong vertical wind shear (“VWS”) in the cloud layer above a shallow boundary layer. The resulting turbulent eddy circulation, reflected as elevated RUC TKE index values, corresponded to the strong wind gusts that were observed by several INL mesonet stations during the afternoon of 19 December.

### 3.2 Case 2: 9 January 2008

During the afternoon of 9 January 2008, high winds were observed at several INL mesonet stations. The highest wind gust of 36 knots was recorded by Aberdeen and Minidoka mesonet stations between 1900 and 2015 UTC (1200 and 1315 MST), and was significantly higher than winds observed over the INL during the same time period. The midday (1900 UTC) RUC three-hour forecast boundary layer turbulence product indicated a maximum or “ridge” in index values over extreme southern Idaho, south of the INL. The combination of strong vertical wind shear across the top of the boundary layer and moderate solar heating of the surface resulted in the generation of turbulent eddy circulations and the high wind gusts observed at Aberdeen and Minidoka.

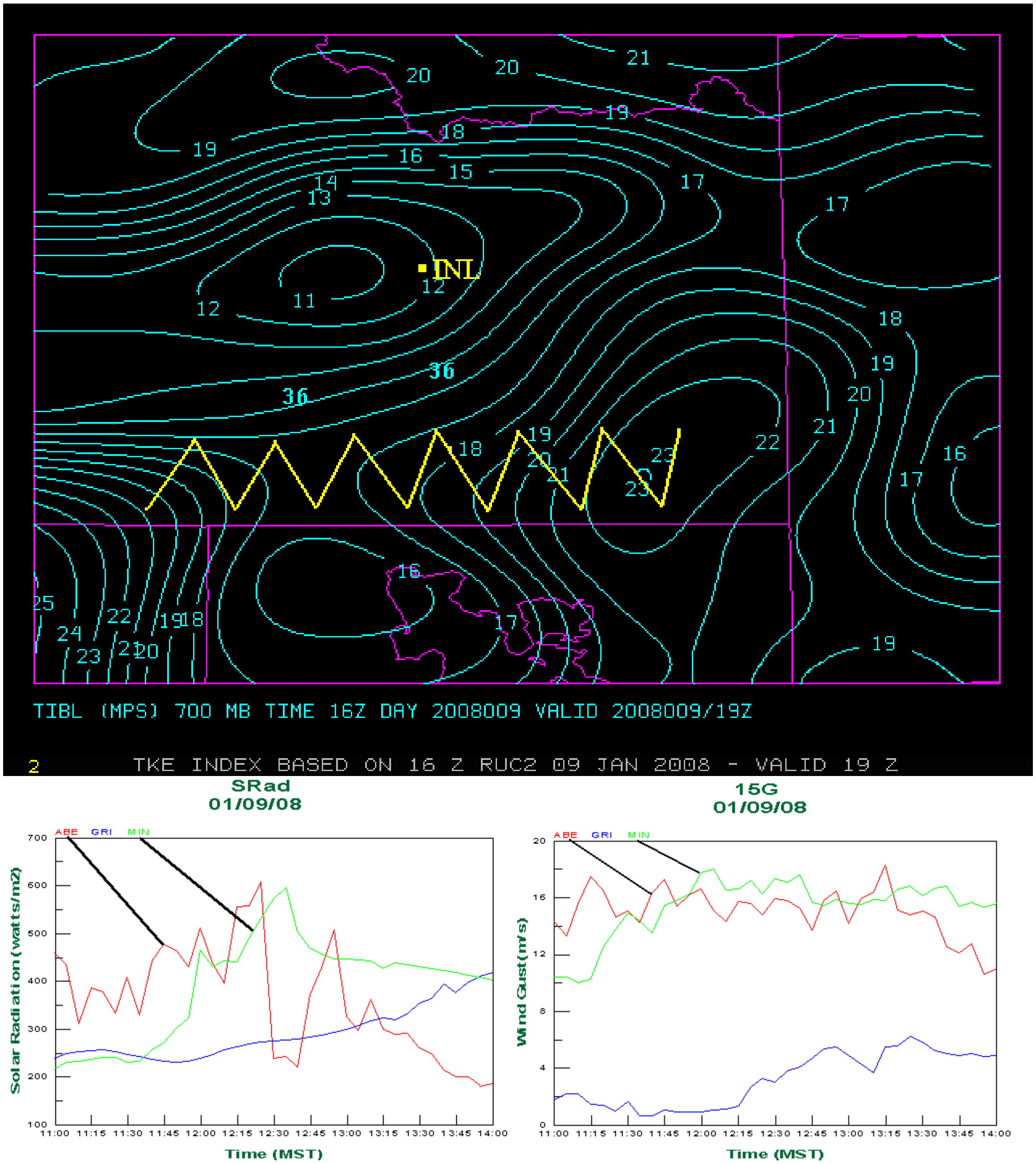

Figure 2. A Turbulent Kinetic Energy (TKE) Index product image (top) based on the Rapid Update Cycle (RUC) model valid at 1900 UTC 9 January 2008 and INL mesonet metograms displaying wind gust (15G, bottom right) and solar radiation (SRad, bottom left) observations between 1800 and 2100 UTC (1100 and 1400 MST).

The TKE image in Figure 2 displays a maximum in index values over extreme southern Idaho, south of the INL. Peak wind gusts of 18 m/s (36 knots), observed at Aberdeen (ABE) and Minidoka (MIN) mesonet stations and plotted in the TKE image, were significantly higher than wind gusts observed over the INL (GRI) at during the same time period. The solar radiation meteogram displayed much higher values over Aberdeen and Minidoka. Thus, the combination of more intense solar heating and stronger vertical wind shear in the boundary layer resulted in turbulent eddy circulations and the strong wind gusts observed over southern Idaho.

### 3.3 Case 3: 15 January 2008

During the morning of 15 January 2008, high winds were observed at several INL mesonet stations in association with a cold front passage. The highest wind gust of 58 knots was recorded by Fort Hall mesonet station at 1120 UTC (0420 MST), and was significantly higher than winds observed over the INL during the same time period. High wind gusts of 40 to 46 knots were then observed over the INL between 1220 and 1250 UTC (0520 and 0550 MST). Morning (1100 to 1200 UTC) RUC three-hour forecast boundary layer turbulence products indicated that a maximum or “ridge” in index values tracked over INL during the time of peak winds. Strong vertical wind shear resulted in the generation of turbulent eddy circulations and high surface wind gusts due to downward momentum transport.

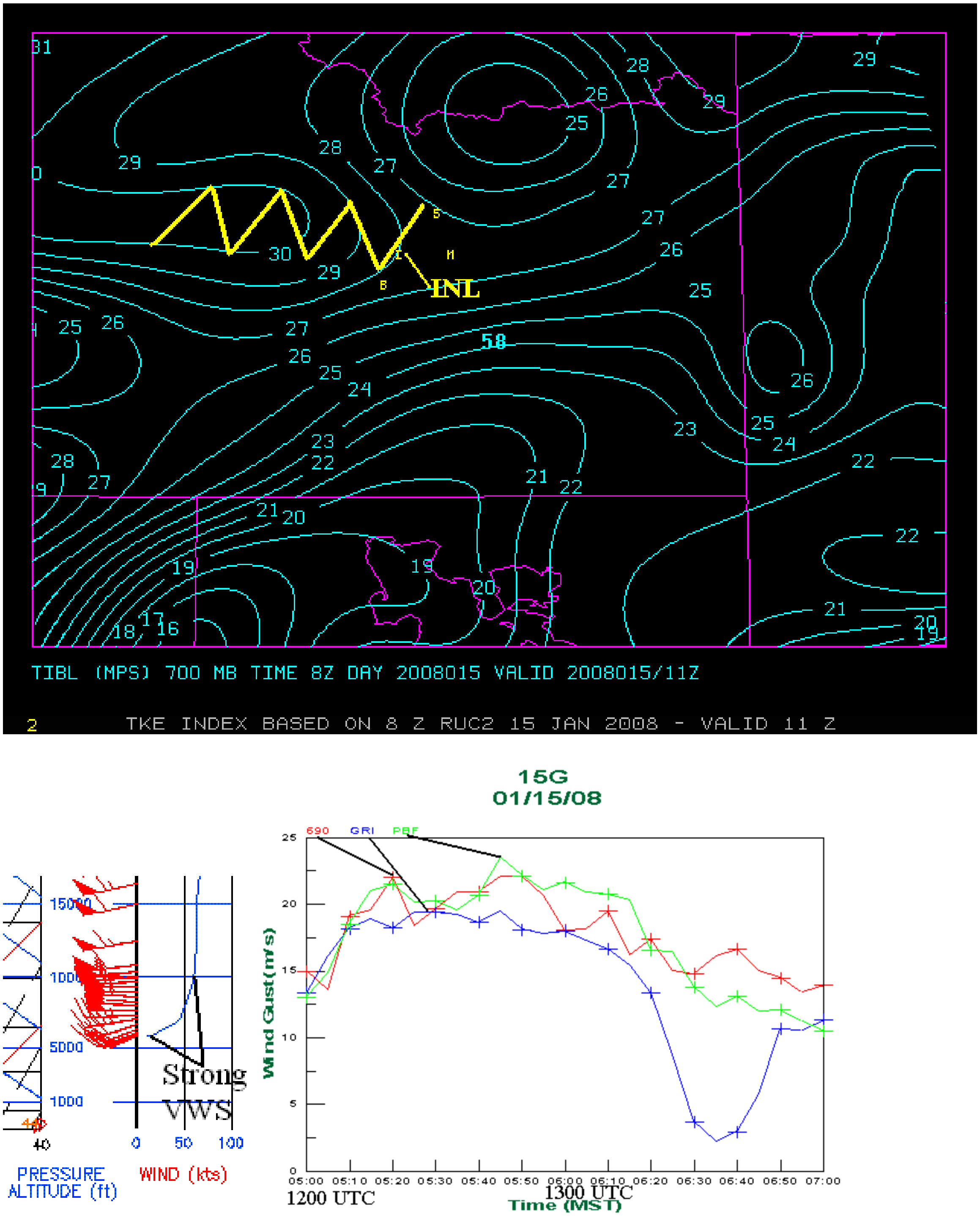

Figure 3. Turbulent Kinetic Energy (TKE) Index product image (top) and vertical wind profile (bottom left) based on the Rapid Update Cycle (RUC) model valid at 1100 UTC 15 January 2008 and an INL mesonet meteogram displaying wind gust (15G, bottom right) observations between 1200 and 1400 UTC (0500 and 0700 MST).

The TKE image in Figure 3 displays a maximum in index values over and west of the INL. The corresponding wind profile over INL in Figure 3 indicated strong vertical wind shear up to 10000 feet above mean sea level (MSL). Winds at 7000 feet above

MSL (2000 feet above ground level) were near 50 knots. Peak wind gusts of 20 to 23 m/s (40 to 46 knots) were recorded by INL mesonet stations 690, GRI, and PBF between 1220 and 1250 UTC. A peak wind gust of 58 knots was recorded about one hour earlier at Fort Hall mesonet station (plotted in image). Thus, strong vertical wind shear in the boundary layer resulted in turbulent eddy circulations and the strong wind gusts observed over southeastern Idaho.

### 3.4 Case 4: 13 February 2008

During the evening of 13 February 2008, high winds were observed at several INL mesonet stations in association with a cold front passage. The highest wind gust of 58 knots was recorded by Dubois mesonet station at 0605 UTC 14 February 2008 (2305 MST 13 February). 0500 to 0600 UTC Rapid Update Cycle (RUC) three-hour forecast boundary layer turbulence products indicated the presence of a maximum or “ridge” in index values over the northern INL mesonet domain near the Montana border. Strong wind shear in an elevated mixed layer resulted in the generation of turbulent eddy circulations and high surface wind gusts due to downward momentum transport.

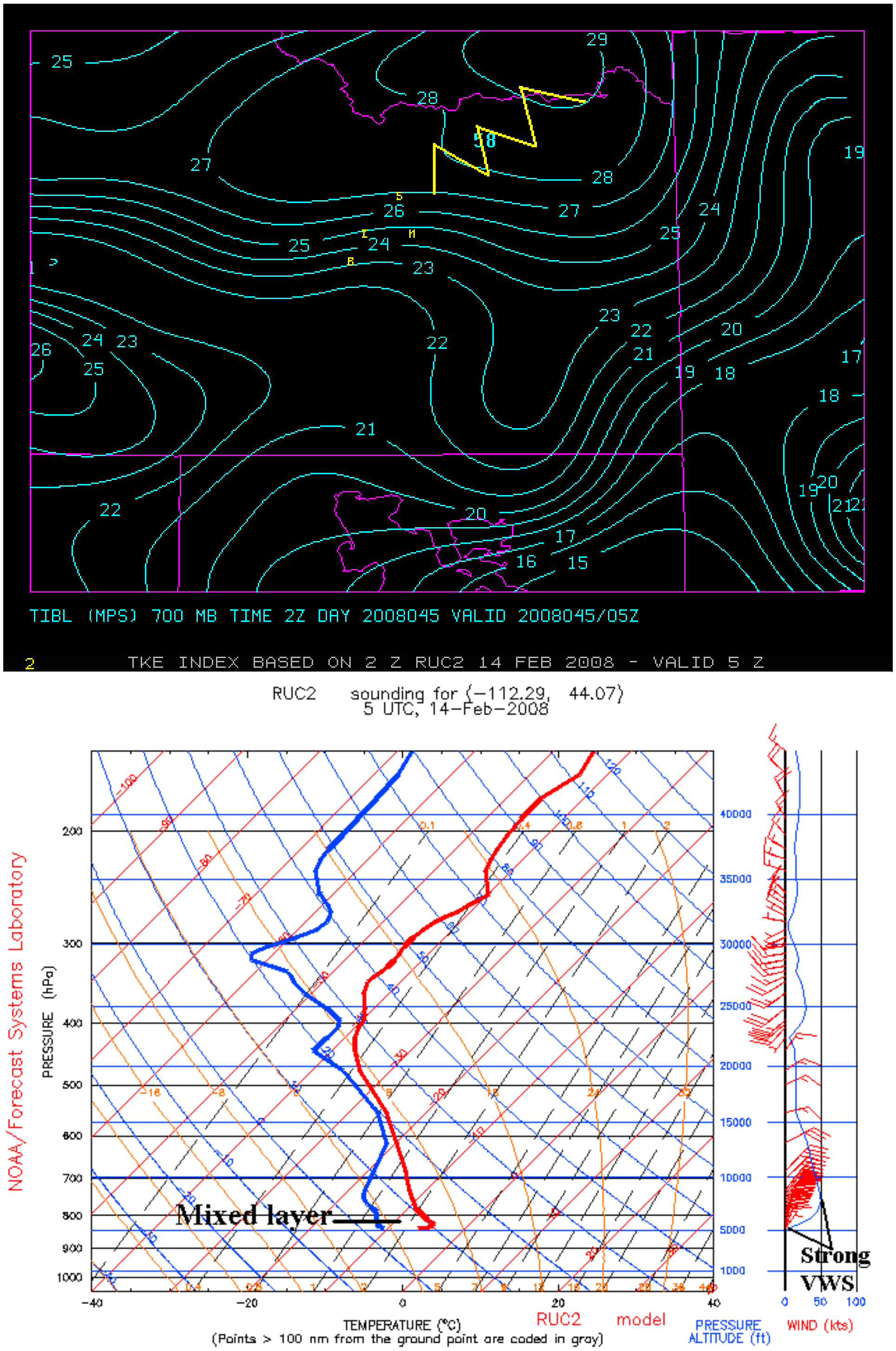

Figure 4. Turbulent Kinetic Energy (TKE) Index product image (top) based on the Rapid Update Cycle (RUC) model valid at 0500 UTC February 14, 2008, a corresponding RUC sounding at Dubois mesonet station (bottom).

The TKE image in Figure 4 displayed a maximum in index values northeast of the INL near the Montana border. The RUC profile in Figure 4 indicated strong vertical wind shear up to 3000 feet above ground level (AGL). Winds at 3000 feet AGL were near 50 knots. A peak wind gust of 58 knots (plotted in image) was recorded at Dubois station at 0605 UTC. Thus, strong vertical wind shear in the boundary layer resulted in turbulent eddy circulations and the strong wind gusts observed over southeastern Idaho.

## 4. Discussion

Analysis of covariance has indicated a strong positive functional relationship between TKE index values and wind gust magnitude over the INL. Correlation (r) was determined to be near 0.79 with a near 100 percent confidence level that this correlation represents a physical relationship between TKE and wind gust magnitude. A *t* test was conducted to determine statistical significance in which a significance level of .01 and a critical value of 2.718 were selected. Comparing the calculated *t* value of 12.64 to the selected critical value confirmed the high statistical significance and confidence level of the correlation between TKE values and observed wind gusts. Thus, comparison of observations of high wind gusts to proximity TKE values has proven to be an effective means to assess the performance of the boundary layer turbulence product. Comparing TKE product images to proximity RUC soundings and radar wind profiles over the INL verified the physical mechanisms (i.e. convection, vertical wind shear) associated with boundary layer turbulence in each case. This validation effort is based on the premise that wind gusts observed at the surface are a manifestation of turbulence generated in the boundary layer.

The case studies also demonstrate the effectiveness of the TKE product in the short-term prediction of boundary layer turbulence and gusty surface winds for various synoptic situations. In two of the four cases, a cold front passage was associated with strong and gusty surface winds. Three-hour turbulence forecasts derived from RUC model data effectively indicated conditions favorable for non-convective, high surface wind gusts. Other cases were associated with intense vertical mixing resulting from a combination of surface heating and strong vertical wind shear. Table 1 lists 12 significant wind events included in the computation of correlation statistics.

## 5. Summary and Conclusions

A boundary layer turbulence index (TIBL) product has been developed to assess the potential for turbulence in the lower troposphere, generated using RUC-2 numerical model data. The index algorithm approximates boundary layer turbulent kinetic energy by parameterizing vertical wind shear, responsible for mechanical production of TKE, and kinematic heat flux, parameterized by the vertical temperature lapse rate $\partial T/\partial z$ and responsible for buoyant production of TKE. Validation for the TIBL product has been conducted for selected nonconvective wind events during the 2008 winter season over the INL mesonet domain. This paper presented studies of four significant wind events between December 2007 and February 2008 over southeastern Idaho. Based on the favorable results highlighted from validation statistics and in the case studies, the RUC

TIBL product should have operational utility in assessing turbulence hazards to low-flying aircraft and ground transportation, and in the assessment of wildfire threat.

| | | | | |
|---|---|---|---|---|
| **RUC TKE** | **2007-2008** | | | |
| **INL Mesonet** | | | | |
| **Correlation (r):** | | | | |
| TKE to measured wind: | | 0.7849 | | |
| | | | | |
| No. of events: | 12. | Mean Wind Speed: | | 45.33 |
| | | Mean TKE: | | 23.75 |
| | | **Measured Wind** | | **RUC** |
| **Date** | **Time (UTC)** | **Speed kt** | **Location** | **TKE** |
| **19-Dec-07** | 2200 | 36 | EBR | 21 |
| | 2200 | 40 | LOS | 22 |
| | 2210 | 41 | PBF | 22 |
| **9-Jan-08** | 1900 | 36 | MIN | 15 |
| | 2010 | 36 | ABE | 15.5 |
| **15-Jan-08** | 1220 | 45 | 690 | 28.5 |
| | 1230 | 39 | GRI | 25.5 |
| | 1250 | 46 | PBF | 26 |
| | 2150 | 55 | DUB | 26.5 |
| **7-Feb-08** | 2300 | 60 | FOR | 29 |
| | 2330 | 52 | MIN | 26 |
| **14-Feb-08** | 600 | 58 | DUB | 28 |

Table 1. Non-convective high wind events as recorded by INL mesonet stations December 2007 and February 2008.